# Sensitivity enhancement using chirp transmission for an ultrasound arthroscopic probe

B. Pialot, A. Bernard, H. Liebgott and F. Varray

*Abstract*— Meniscal tear in the knee joint is a highly common injury that can require an ablation. However, the success rate of meniscectomy is highly impacted by difficulties in estimating the thin vascularization of the meniscus, which determines the healing capacities of the patient. Indeed, the vascularization is estimated using arthroscopic cameras that lack of a high sensitivity to blood flow. Here, we propose an ultrasound method for estimating the density of vascularization in the meniscus during surgery. This approach uses an arthroscopic probe driven by ultrafast sequences. To enhance the sensitivity of the method, we propose to use a chirp-coded excitation combined to a mismatched compression filter robust to the attenuation. This chirp approach was compared to a standard ultrafast emission and a Hadamard-coded emission using a flow phantom. The mismatched filter was also compared to a matched filter. Results show that, for a velocity of a few mm.s$^{-1}$, the mismatched filter gives a 4.4 to 10.4 dB increase of the signal-to-noise ratio compared to the Hadamard emission and a 3.1 to 6.6 dB increase compared to the matched filter. Such increases are obtained for a loss of axial resolution of 13% when comparing the point spread functions of the mismatched and matched filters. Hence, the mismatched filter allows increasing significantly the probe capacity to detect slow flows at the cost of a small loss in axial resolution. This preliminary study is the first step toward an ultrasensitive ultrasound arthroscopic probe able to assist the surgeon during meniscectomy.

## I. Introduction

M ENISCAL tears are the most common injuries treated in the knee joint [1]. During a knee arthroscopy, the choice to perform a meniscectomy is directly linked to the density of vascularization in the tear zone. If the vascularization is too poor, the damaged zone is removed but at the risk of future complications [2]. Currently, surgeons can use fiber optic cameras for estimating the density of vascularization. But, these cameras cannot clearly separate the vascularization from the surrounding tissue, thus resulting in a poor sensitivity and contributing to the low success rate of meniscectomy [3].

With ultrafast sequences and Singular Value Decomposition (SVD) filtering, ultrasound is now capable of efficiently discriminating slow blood flows from surrounding tissues [4]. This improvement has introduced ultrafast ultrasound imaging as a new modality for micro-vascularization imaging [5]. If the signal of the targeted vessels is too weak, ultrasound backscattering can be enhanced by injecting contrast agents to the imaging subject. An alternative that does not require any injection or that can be combined with contrast agents is to use coded emissions [6]. In particular, the use of Hadamard-coded sequences has been shown to enhance the signal-to-noise ratio (SNR) in ultrafast sequences [7]. We can also mention techniques that exploit the coherence of backscattered ultrasound signal [8].

Prior to ultrafast imaging, chirp excitations have been demonstrated to significantly increase ultrasound imaging capabilities [9]. Since, chirp imaging have been successfully used in various applications [10]–[13] but very marginally for ultrafast flow imaging despite its advantages. When receiving a chirp signal, a compression filter that allows recovering a point spread function (PSF) close to conventional pulse imaging is applied. The most common compression method is the matched filter that simply consists in the convolution of the received signal with a time-reversed copy of the emitted chirp [6], [14]. The matched filter can also be weighted by a temporal window in order to achieve a compressed pulse with lower side-lobes but at the cost of reduced SNR and axial resolution [6]. Another compression approach is the Wiener filter that is derived from linear least squares estimation [15]. We can also cite the Fractional Fourier Transform that can be used when chirp signals overlap because of thin imaging layers [16].

Ultrasound attenuation have to be taken into account for an optimal chirp compression when imaging the human body. Indeed, the strong attenuation of human tissues shifts down the

This work was supported by the FEDER project MenisCare. It was also performed within the framework of the LABEX PRIMES (ANR-11-LABX-0063) of the Université de Lyon, within the programme 'Investissements d'Avenir' (ANR-11-IDEX-0007), operated by the French National Research Agency (ANR). This material is based upon work done on the PILoT facility (PILoT, INSA-Lyon). The RF Verasonics generator was cofounded by the FEDER program, Saint-Etienne Metropole (SME) and Conseil General de la Loire (CG42) within the framework of the SonoCardio-Protection Project leaded by Pr Pierre Croisille.

All authors are with the Univ Lyon, INSA-Lyon, Université Claude Bernard Lyon 1, CNRS, Inserm, CREATIS UMR 5220, U1294, F-69621, Lyon, France (e-mails : baptiste.pialot@creatis.insa-lyon.fr, adeline.bernard@creatis.insa-lyon.fr, liebgott@creatis.insa-lyon.fr, francois.varray@creatis.insa-lyon.fr).

peak frequency (i.e. the frequency for which the spectral amplitude is maximum) and the bandwidth of the received signal along the depth [12], [17]. Thus, the matched filter is no more adapted to the received signal. To tackle this effect, the matched filter can be adapted to the received signal instead of being a direct copy of the transmitted chirp. In the spectral domain, it consists to shift the central frequency of the filter and to reduce its bandwidth in order to fit with the received spectrum. By doing this adaptation, the matched filter becomes a mismatched filter and the SNR gain of the compression can be increased [12]. Another approach can be to directly enhance the attenuated frequencies when emitting the chirp signal [18].

In this experimental study, we introduce a new method for ultrasound imaging of the meniscus vascularization during surgery using an arthroscopic probe. In particular, we investigate on a phantom how to increase the sensitivity of the probe using ultrafast chirp-coded emission. Because the ultrasonic signal of the meniscus vascularization should be very weak, we chose specific experimental conditions in order to mimic a poor SNR imaging environment. To compensate the attenuation of the phantom, we apodize the matched filter (mismatched filter) by a Gaussian window to adapt the compression filter to the received spectra. This mismatched filter allows selecting the frequencies for which the power Doppler signal is the less attenuated. Finally, we demonstrate that chirp-coded plane wave emissions combined to the mismatched filter allows visualizing slow flows with a higher sensitivity than the Hadamard-coded emission.

The paper is organized as follows: we first describe the imaging system, experimental set-up and acquisition sequences as well as the compression filters. We then present and discuss the obtained results before giving a short conclusion on the outcomes of our work.

## II. MATERIAL AND METHODS

*A. Imaging system*

An arthroscopic linear probe with a central frequency of 11.6 MHz was specifically designed and manufactured (Vermon, France). The probe consists in a 22-cm-long stainless steel rod with a diameter of 4 mm. Sixty-four piezoelectric elements are disposed on the side of the rod, close to its tip. The elements are aligned with a pitch of 0.20 mm, corresponding to 1.5 wavelengths at a speed of sound of 1540 m/s. A circular handle encompasses the bottom of the probe to facilitate its manipulation. The probe has been conceived to be inserted in the knee during meniscus surgery such as the cameras mentioned in the introduction.

A Verasonics Vantage scanner was used to drive the probe and to acquire RF data at a sampling frequency of 50 MHz. The driving voltage from the scanner to the probe was set to 15 V peak-to-peak during all experiments. Coded emissions were generated using the tri-state pulser (ArbWave Toolbox in Verasonics) provided by the Vantage scanner. All post-processing and images formation steps were performed in Matlab (Mathworks, Natick, MA, USA).

*B. Experimental phantom*

The following procedure was performed two times and consequently the results presented in the results section come from two different experimental phantoms. The two phantoms will be referred to as phantom A and phantom B. The phantom A was used to evaluate the compression method for the chirp acquisitions. The phantom B was used to compare the different emission approaches.

The phantoms consisted in a tissue-mimicking layer surrounding a silicon tube (inner diameter: 0.5mm, outer diameter: 1mm). The tissue-mimicking layer was composed of a mixture of water (95% of its mass), agar (4%) and silica powder (1%, 10 $\mu$m average diameter). The ultrasound attenuation coefficient of the tissue-mimicking layer was measured to 0.17$\pm$0.03 dB.cm$^{-1}$.MHz$^{-1}$ using a substitution method [19]. The silicon tube was first glued through a small plastic box with a thin layer of ultrasound gel on its surface to ensure its coupling with the agar mixture. A hole was perforated in the plastic box approximatively 2 cm away from the tube to pass the probe in the box during experiments. Then, the mixture of agar was mixed and melted at 80°C before being carefully pounded around the tube in the plastic box. Finally, the probe was inserted directly in the tissue-mimicking layer formed by the agar mixture through the hole in the plastic box. For ultrasound acquisitions, a flow of cellulose powder (10 $\mu$m average diameter) in water was imposed in the tube by a syringe pump (InfusionOne, New Era Pump Systems, New-York, USA). Fig. 1 shows a picture of the probe and a scheme of the experimental set-up.

*C. Ultrasound sequences and post-processing*

All ultrasound sequences were based on plane waves emissions followed by coherent compounding after delay-and-sum beamforming of the RF data [20]. All sequences were fired one after another while the cellulose powder in the syringe was continuously mixed using a magnetic bar. For each sequence, 4 plane waves where fired at angles of -0.5°, 0.5°, -1° and 1° and at a PRF of 2000Hz thus allowing an effective PRF of 500Hz between each compounded frame. The small variations between angles were chosen because of the desired low SNR as well as to keep an optimal field of view. A total of 128 compounded frames were acquired for all sequences and were then clutter filtered with SVD before being incoherently summed to form a power Doppler image.

Three different types of ultrasound sequences were used: a basic plane waves sequence (3 cycles), a multiplane waves sequence (also 3 cycles) with a 4 angles Hadamard basis encoding [7] and a chirp-coded plane waves sequence. The chirp sequence was performed using a linear chirp such as in the study of Mamou *et al.* [10]:

$$s(t) = w(t)\cos\left(2\pi f_1 t + \pi\frac{f_2 - f_1}{T}t^2\right) \quad (1)$$

where $t$ is the time vector, $f$ is the chirp central frequency, $f_1$ and $f_2$ are the upper and lower chirp frequencies, $T$ is the chirp duration and $w(t)$ is an apodization window to reduce frequency ripples. Experimental values for the chirp parameters were 3 μs for $T$, 5 MHz for $f_1$ and 18 MHz for $f_2$. Values for $f_1$ and $f_2$ were chosen accordingly to the measured -6 dB bandwidth of the probe. The apodization of the chirp was performed with a Tukey window presenting a 15% cosine fraction. The chirp of equation (1) was inputted at a sampling frequency of 250 MHz to the tri-state pulser of the Vantage scanner.

*D. Decoding of signals*

The decoding of coded ultrasound sequences was performed individually for each acquisition channel on raw RF data. Hadamard sequences were decoded by multiplications of received RF signals following the Hadamard basis as described in the study of Tiran *et al.* [7]. For chirp sequences, the matched filter and the mismatched filter were compared.

*1) Matched filter*

The matched filter compression of RF signals was performed in the frequency domain as follows:

$$\widehat{RF} = \mathfrak{I}^{-1}(\mathfrak{I}(RF) \circ \mathfrak{I}(S)^*) \quad (2)$$

where $\mathfrak{I}(.)$ and $\mathfrak{I}^{-1}(.)$ are the Fourier transform and inverse Fourier transform, $\circ$ is the element-wise product, $*$ is the complex conjugate, $RF$ is a vector containing the received $N$ samples, $\widehat{RF}$ is a vector with the $N$ compressed samples and $S$ is a vector containing $L < N$ samples of $s(t)$ in equation (1). The Fourier transforms $\mathfrak{I}(RF)$ and $\mathfrak{I}(S)$ of the temporal vectors $RF$ and $S$ were estimated on 1024 samples.

*2) Mismatched filter*

As stated in the introduction, Ramalli *et al.* have demonstrated how to incorporate attenuation into chirp compression using a mismatched filter [12]. In their study, the central frequency of the mismatched filter was first estimated from the spectrum of the received echoes. Then, the -6 dB bandwidth of the filter was deduced from the shift under the assumption of a linear law between attenuation and frequency. Here, we chose to use a similar approach to incorporate the attenuation of the phantom in the compression process.

First, the peak frequency $f_m$ of the received spectrum averaged on the 128 frames, 4 compounding angles and 64 receiving elements was estimated in a ROI of 6 mm (200 samples) surrounding the measurement tube. The RF signal corresponding to the ROI was zero-padded to obtain 1024 frequency points for the estimation of $f_m$. Then, the following Gaussian function normalized between 0 and 1 was computed:

$$G(f) = e^{\frac{-(f-f_m)}{\sqrt{2}\sigma}} \quad (3)$$

where $\sigma$ is a parameter that rules the bandwidth of the function.

The Gaussian function of equation (3) was used to apodize the matched filter $\mathfrak{I}(S)^*$. This apodization led to the following compressed RF signal:

$$\widehat{RF} = \mathfrak{I}^{-1}(\mathfrak{I}(RF) \circ [G \circ \mathfrak{I}(S)]^*) \quad (4)$$

where $[G \circ \mathfrak{I}(S)]$ is the mismatched filter given by the apodization and $G$ is a vector that contains $N$ samples of the Gaussian function of equation (3).

The main idea of the Gaussian apodization is to select the less attenuated frequencies of the chirp signal to perform the compression. The compromise between the gain in SNR and the loss of axial resolution is effectively ruled by the parameter $\sigma$. An increase of $\sigma$ will broaden the bandwidth of the received spectrum used for the compression, thus resulting in an enhancement of the axial resolution and a decrease of the SNR. On the contrary, if the compressed bandwidth is restrained to its most energetic part by lowering $\sigma$, the SNR will be increased but the axial resolution will be degraded.

*E. Safety considerations*

Chirp-coded emissions require to transmit longer signal in the imaging medium than conventional pulse imaging. Thus, it is important to evaluate the safety of chirp sequences when targeting an *in vivo* application. Here, we chose to measure the Mechanical Index (MI) as well as the Intensity Spatial Peak Temporal Average (ISPTA) [21]. The acoustic pressure field radiated by the probe was measured in water with a hydrophone and the indices were computed as follows:

$$ISPTA = PRF \int_0^T I dt \quad (5)$$

$$MI = \frac{p_r}{\sqrt{f_e}} \quad (6)$$

where $I$ is the intensity of the acoustic field at the point where the peak acoustic pressure is maximum, $p_r$ is the peak rarefication pressure at the same point, $f_e$ is the emission frequency, $PRF$ is the pulse repetition frequency between compounding angles and $T$ is the duration of the chirp.

The MI is related to the potential risk of biomedical unwanted effects such as cavitation. Its value should theoretically stay the same between a pulse emission and a chirp-coded emission. The ISPTA is a measure of the averaged intensity that is send to the imaging medium during the sonication and is related to tissue heating [21]. For endoscopic imaging, the Food and Drug Administration (FDA) recommends to not exceed an ISPTA of 94 mW.cm$^{-2}$ in water. The MI recommended by the FDA in water for general ultrasound imaging is 1.9.

*F. Quality metrics*

*1) Signal and contrast*

The efficiency of the emission schemes and of the compression methods was quantified in terms of signal-to-noise ratio (SNR) and contrast-to-noise ratio (CNR) on power Doppler images. The zone corresponding to the measurement tube in the images was segmented to obtain the signal ROI $I_s$ and to compute its mean $\mu_s$ and standard deviation $\sigma_s$. The signal ROI was then translated to 4 mm down the image to obtain a noise ROI $I_n$ with its mean $\mu_n$ and standard deviation $\sigma_n$. SNR and CNR were finally computed as follows [8], [22]:

$$SNR = 20 \log \left( \frac{\sqrt{\frac{1}{N}\sum_{i=1}^{N} I_s^{\,2}(i)}}{\sqrt{\frac{1}{N}\sum_{i=1}^{N} I_n^{\,2}(i)}} \right) \quad (7)$$

$$CNR = 20 \log \left( \frac{|\mu_s - \mu_n|}{\sqrt{0.5(\sigma_s^{\,2} + \sigma_n^{\,2})}} \right) \quad (8)$$

The SNR quantifies the level of the signal ROI compared to the noise ROI. The CNR quantifies the detection of the signal ROI compared to the noise ROI by taking account of the contrast between them.

*2) Imaging resolution*

The Full Width at Half Maximum (FWHM) of the PSF was measured in axial and lateral directions for each sequence. This measure was performed on raw compounded images acquired on a wire phantom. The wire phantom was disposed in the transversal direction of the probe to mimic a point scatterer.

### III. RESULTS

The measured MI for the driving voltage of 15 V peak-to-peak was 0.34 and the measured ISPTA was 0.12 mW.cm$^{-2}$. Both of these values are well below the maximum values of 1.9 and 94 mW.cm$^{-2}$ recommended by the FDA in water.

Fig. 2A shows the evolution of the peak frequency of the received raw RF spectrum as a function of depth. The spectrum comes from phantom A imaged at a mean flow velocity of 10 cm.s$^{-1}$. For each depth, the spectrum has been estimated in a window of 6 mm corresponding to 200 samples. The spectrum has been furthermore averaged on all frames, compounding angles and receiving elements. It can be seen that the peak frequency is at 12.3 MHz for a depth of 0.3 cm and then drops abruptly to 7.3 MHz for a depth around 0.5 cm. Once this drop is reached, the peak frequency slowly decays to a value of 6.8 MHz for an imaging depth of 2.7 cm.

Fig. 2B shows the averaged raw RF spectra used for computing the peak frequencies at different imaging depths in Fig. 2A. It can be seen that the bandwidth of the spectra reduces significantly with the depth. As already seen in Fig. 2A, the spectrum at a depth of 0.3 cm have a peak frequency of 12.3 MHz. All the others spectra, from an imaging depth of 0.6 cm to 2.5 cm, have a peak frequency between 6.8 MHz and 7.3 MHz.

Fig. 3A reports the SNR and the CNR measured for a chirp power Doppler image acquired in phantom A at a mean flow velocity of 10 cm.s$^{-1}$. Both indicators are shown as a function of the parameter $\sigma$ used for the mismatched filter and for the matched filter. The SNR decreases almost linearly with $\sigma$ from 20.0 dB if $\sigma = 2$ MHz to 13.9 dB if $\sigma = 6$ MHz. The CNR also decreases with $\sigma$ but in a less pronounced way from a value of 8.7 dB to a value of 8.3 dB. In comparison, the SNR and the CNR for the matched filter are 10.3 dB and 8.1 dB, respectively.

Fig. 3B shows the resolution measurement corresponding to each SNR and CNR values in Fig. 3A. The axial resolution obtained with the mismatched filter decreases with $\sigma$ and finally converges to the axial resolution of 260 $\mu$m provided by the matched filter. The worst axial resolution, corresponding to a SNR of 20.0 dB, is 334 $\mu$m and is obtained for $\sigma = 2$ MHz. For $\sigma = 5$ MHz, the axial resolution of the two filter are almost identical while the SNR of the mismatched filter is still 3.6 dB above the SNR of the matched filter. On the contrary, the lateral resolution is strongly enhanced if the mismatched filter is used instead of the matched filter. Indeed, the lateral resolution for the mismatched filter stays around a value of 434 $\mu$m if $\sigma$ is between 2 and 5 MHz. In comparison, the lateral resolution provided by the matched filter is 568 $\mu$m. The lateral resolution of the mismatched filter becomes close to the one for the matched filter only if $\sigma = 6$ MHz, with a value of 548 $\mu$m. In overall, a good compromise is obtained between the SNR gain and the imaging resolution when $\sigma = 2.5$ MHz. In that case, the SNR gain is 8.6 dB while the axial resolution increases by only 35 $\mu$m (+13%).

Fig. 4A reports the matched and the mismatched filter in the temporal domain. The mismatched filter is shown for $\sigma = 2$ MHz and $\sigma = 5$ MHz. It can be seen that the mismatched filter preserves all the frequencies of the original deconvolution chirp when $\sigma = 5$ MHz. On the contrary, a significant part of the highest frequencies is totally damped at $\sigma = 5$ MHz.

Fig. 4B shows the matched and the mismatched filter as in fig. 4A but in the frequency domain. The dynamic range is between 0 dB and -30 dB. The matched filter have an almost flat amplitude between 6 MHz and 18 MHz. In contrast, both mismatched filters present a strongly reduced bandwidth with a peak frequency around 6.7 MHz. The influence of the parameter $\sigma$ is highlighted as it can be seen that the mismatched filter has a much wider bandwidth when $\sigma = 5$ MHz.

Fig. 4C shows the averaged raw RF spectrum as acquired by the probe in the tube and also after a compression with the matched filter, the mismatched filter with $\sigma = 2$ MHz and the mismatched filter with $\sigma = 5$ MHz. The dynamic range is again between 0 dB and -30 dB. First, it can be seen that the received spectrum is subjected to a strong attenuation, which distorts its shape. Indeed, we can observe a reduction of the spectrum bandwidth and a decrease of its peak frequency to a value around 6.7 MHz. Therefore, the matched filter in Fig. 4B does not fit anymore with the received spectrum. For the compressed spectra, it can be seen that both mismatched filters only keep the most energetic part i.e. the low frequency components of the received spectrum. The spectra compressed by the two mismatched filters have approximatively the same -6 dB bandwidth, between 6.3 MHz and 7.8 MHz. The main difference between the two filters is at which frequency the compressed spectrum reaches the maximum extent of its -30 dB bandwidth. Indeed, if $\sigma = 5$ MHz, the compressed -30 dB bandwidth extends to 12 MHz but only to approximatively 10 MHz if $\sigma = 2$ MHz. For the spectrum compressed with the matched filter, the -6 dB bandwidth is close to the ones of the two mismatched filters. However, the rest of the spectrum presents a much higher and homogeneous amplitude than with a mismatched filter, even at frequencies for which the received spectrum is strongly attenuated. For example, the amplitude of the spectrum compressed by the matched filter is still above -15 dB for a frequency of 16 MHz whereas the received spectrum amplitude is at -25 dB for this frequency. Also, it can be seen that the spectrum compressed with the matched filter presents more local fluctuations of its amplitude that with the mismatched filter. This difference indicates that the compressed spectrum is more corrupted by noise when the matched filter is used.

Fig. 5 reports power Doppler images obtained in phantom B with a mean flow velocity ranging from 0.85 mm.s$^{-1}$ to 4.2 mm.s$^{-1}$. The images are shown for the plane waves and Hadamard sequences as well as for the chirp sequences. For the chirp sequences, compressions using the matched filter and a mismatched filter with $\sigma = 2.5$ MHz are both reported. It can be seen that the signal coming from the flow cannot be properly imaged using only plane waves and this for the whole velocity range. The Hadamard sequence allows visualizing the flow if the velocity is above 3.4 mm.s$^{-1}$ but below this value, the flow is entirely restituted only if chirp sequences are used. For those chirp sequences, it can be seen that the mismatched filter reduces the background noise in comparison to the matched filter and this for the whole velocity range. In particular, the flow is barely detectable with the eye at 1.7 mm.s$^{-1}$ using the matched filter but appears clearly using the mismatched filter. Also, it can be seen that several images present a strong electronic noise in their sides. This effect is cancelled when the mismatched filter is used as well as the Hadamard-coded emission.

Fig. 6A reports SNR measurements in Power Doppler images from Fig. 5. All the sequences fail to provide a strictly positive SNR at the bottom velocity of 0.85 mm.s$^{-1}$. For the plane waves sequence, the SNR stays negative in the whole velocity range thus confirming that the flow cannot be detected using this approach. The Hadamard sequence slightly raises the SNR above 0 when the mean velocity is above or equal to 3.4 mm.s$^{-1}$, which makes the flow detectable in this range as observed in Fig. 5. For chirp sequences, the SNR presents the same behavior for all compression filters with an increase until a velocity of 4.2 mm.s$^{-1}$ followed by a slight decrease. Based on an extrapolation of the behavior demonstrated by the curve, the SNR measured for the matched filter seems to cross 3 dB for a mean velocity around 3 mm.s$^{-1}$. For the mismatched filter, the 3 dB mark should be reached for a velocity around 1.5 mm.s$^{-1}$. For the bottom detectable velocity of 1.7 mm.s$^{-1}$, the SNR is 1.2 dB and 4.3 dB for the matched filter and the mismatched filter respectively. At the top velocity of 4.2 mm.s$^{-1}$ the SNR is 4.2 dB for the matched filter and 10.8 dB for the mismatched filter. Overall, the mismatched filter provides a SNR gain from 4.4 dB to 10.4 dB when compared to the Hadamard emission in the investigated velocity range.

Fig. 6B reports the CNR measured for the same Power Doppler images than in Fig. 6A and Fig. 5. For the sake of clarity, values lower than 0 dB are not shown. It can be seen that the CNR for the mismatched filter is positive for all mean velocities above the bottom value of 0.8 mm.s$^{-1}$. All the others emission schemes have a negative CNR except the matched filter that have a positive CNR when the velocity is above 3.4 mm.s$^{-1}$. At the top velocity of 4.2 mm.s$^{-1}$, the CNR provided by the matched filter is 1.1 dB and the CNR for the mismatched filter is 2.1 dB.

## IV. Discussion

In this study, we have introduced a new arthroscopic probe for imaging the meniscus vascularization during surgery. The probe was driven using ultrasound ultrafast sequences and was tested experimentally on a flow phantom. As the meniscus vascularization should be very thin, we purposely acquired the images at a flow velocity of a few mm.s$^{-1}$ and performed low SNR acquisitions. To enhance the sensitivity of the probe, we proposed to use coded excitations. In particular, we compared Hadamard and chirp-coded sequences and observed that the chirp sequences were able to detect slow flows with a better SNR. In addition, a mismatched filter was used for the compression of chirp signals to account for the effect of attenuation. This filter produced stronger SNR gains than a standard matched filter but degraded the axial resolution of the probe because of a loss of bandwidth during the compression. However, it was demonstrated that the axial resolution could be almost recovered by enlarging the window of the mismatched filter while still producing a significant SNR gain. Also, it has to be noted that the lateral resolution was enhanced by the mismatched filter compared to the matched filter.

Theoretically, the gain of SNR using Hadamard emission with (7) should be equal to $20 \log(N) = 20 \log(4) \approx 12 dB$ where $N$ is the number of emission angles. This gain should normally make the Hadamard-coded sequence competitive with the chirp emission. But, here, the SNR values for Hadamard-coded Power Doppler images are well below this theoretical enhancement. In fact, a factor around 4 was effectively observed for the SNR of B-mode images but not in the SVD-filtered Power Doppler images that are supposed to only display the flow. This can be explained by the fact that the pulsed approach was inherently not sensitive enough to image the flow in the tube, even if a Hadamard summation was performed on received signals. This could maybe be resolved by adding more Hadamard-coded angles at the emission but at the cost of a reduced frame-rate.

As stated in the method section, the use of a compression filter adapted to the attenuated received spectrum was already demonstrated in a past study but only on B-mode images [12]. In particular, the authors also highlighted the link between the SNR gain, the imaging resolution and the apodization window. In the original study, the mismatched filter was a matched filter apodized by a Hamming window. Here, we chose to use a Gaussian window for the apodization because of the quasi-linear relationship that was found between the parameter $\sigma$ and the SNR. It has to be noted that, in theory, the optimal apodization window would probably be the one that follows the attenuation law of the imaging medium. But, such a window would demand a prior knowledge of the attenuation coefficient, which will be complicated in practice.

One of the main limitation of the used mismatched filter is that the shift of the peak frequency is depth-dependent. However, here, the variation of this frequency shift was not very marked beyond a relatively small depth value. Indeed, all the received spectra acquired below a depth of 5 mm had a very similar peak frequency. This result suggests that, even if the ROI used for the application of the mismatched filter is enlarged, it should be still possible to obtain a significant SNR gain over the whole ROI as already observed [12]. However, it is important to note that the attenuation coefficient of our phantom is well below the reported values in cartilage. For instance, Nieminen *et al.* found an attenuation coefficient of 4.0±1.4 dB.cm$^{-1}$.MHz$^{-1}$ in bovine cartilage with n = 10 specimens [23]. Thus, we can expect a stronger shift of the peak frequency during *in vivo* acquisition. In addition, the amplitude of the received spectrum at its peak frequency will be more attenuated. Therefore, the SNR and the imaging resolution will be degraded. However, this degradation can be compensated by a higher emission voltage, in the limits of the recommended values for the MI and the ISPTA. Also, since the frame rate is not critical for our application, we could certainly increase the number of compounding angles and/ or acquire more frames for compensating the SNR loss. In all cases, it has to be noted that a low imaging depth will be used to detect the meniscus vascularization thanks to the arthroscopic process. This low imaging depth is the main aspect that should guarantee the applicability of our probe.

In a medical context, an important aspect of the proposed method should be its capacity to assist the surgeon in real time. However, we deliberately chose to not concentrate this present study on such real-time aspect. Indeed, the computational time will be a function of the imaging parameters such as the number of frames needed to form a power Doppler image. To accurately determine these parameters, an *in vivo* experiment will be necessary. Nonetheless, it has be proven that chirp imaging can be performed in real-time if chirp signals are compressed after beamforming, IQ demodulation and downsampling [13], [14]. We did not use this approach here since we wanted to make a strict comparison between techniques, without downsampling. Future experimentations will evaluate if the compression on downsampled IQ data is sufficient for real-time imaging of very low flows with chirp ultrasound.

During the compression, the Gaussian mismatched filter performs a low-pass filtering of the received chirp signal with a cut-off frequency that is ruled by the parameter $\sigma$. Theoretically, a chirp signal emitted with a bandwidth $B$ and compressed by a matched filter will have a main-lobe width proportional to $\frac{1}{B}$, such as a pulse emitted with the same bandwidth [6]. However, with the Gaussian mismatched filter, the main lobe width will be proportional to $\frac{1}{B'}$ where $B' < B$ is the bandwidth of the compressed and low-pass filtered chirp signal. This broadening of the main lobe will then cause a loss of axial resolution. As explained in [12], the bandwidth of the attenuated chirp signal prior to compression can be approximated as $B = \frac{d}{A.2D}$ where $d$ is the dynamic range (for example -30 dB), $A$ is the attenuation coefficient and $D$ is the imaging depth. To obtain an optimal SNR for a given dynamic range, the user should choose $\sigma$ such that the mismatched filter bandwidth matches $B$. If the goal is to boost the SNR but also to preserve some axial resolution, the users should choose $\sigma$ such that the mismatched filter bandwidth is larger than $B$. The drawback of this approach will be a potential amplification of noisy frequency components outside the dynamic range $d$. Here, we chose to use $\sigma = 2$ MHz and $\sigma = 5$ MHz in Fig. 4 for illustrational purpose and $\sigma = 2.5$ MHz in Fig. 5 and Fig. 6 because this value allows a rather good compromise between the SNR gain and the axial resolution. In practice, $\sigma$ could certainly be determined automatically under the condition of an optimal SNR. Nevertheless, the compromise between SNR and axial resolution will be probably a subjective choice depending on the user. Hence, we prefer to think $\sigma$ as a parameter which will be tunable during the imaging process, depending on the density of vascularization of the meniscus. The interested reader can find additional experimental data and guidelines for chirp compression in an attenuating medium in references [12], [13], [17] and [18].

There are very few quantitative studies about the meniscus vascularization. Indeed, to the author's knowledge, there is no data about the range of blood velocities encountered in this zone of the human body. As stated in the introduction, it has been proven that ultrasound with compounding and SVD filtering can detect low flows in the capillaries of strongly vascularized area such as the brain or the kidney [4]. But, in the meniscus, the backscattered ultrasonic signal by blood might be lower due to a much less present vascularization. Hence, the use of chirp-coded emission with a compression filter robust to the attenuation will be highly valuable when the probe will be used *in vivo*. Nonetheless, an important limit of our study is that the measurements have been performed in a relatively large tube. Indeed, the vascularization of the meniscus can be rather expected to be a group of thinner vessels that backscatter the ultrasonic signal incoherently from one vessel to another. Thus, before *in vivo* experimentations, an additional study could be performed on microfluidic systems to evaluate how the method behaves when a smaller and incoherent structure is investigated. Finally, it has to be noted that photoacoustic could be an interesting alternative to ultrasound for imaging the meniscus vascularization as it provides a better resolution. However, photoacoustic require a specific set-up, more complex than classical ultrasound.

## V. Conclusion

This study introduced a new ultrasound arthroscopic probe for imaging the meniscus vascularization during ongoing surgery. A chirp-coded sequence with a compression filter robust to attenuation has been demonstrated to strongly enhance its sensitivity to very low flows. Next steps will be to optimize the imaging parameters and the compression filter through *in vivo* experimentations. More generally, we hope that this study can help to demonstrate how miniaturized ultrasound probes can be used as surgical tools. In that context, the use of chirp-coded excitation is an interesting approach to make these probes highly sensitive despite their technical limitations.

Acknowledgment

The authors would like to thank Frank Nicolet for insightful discussions.

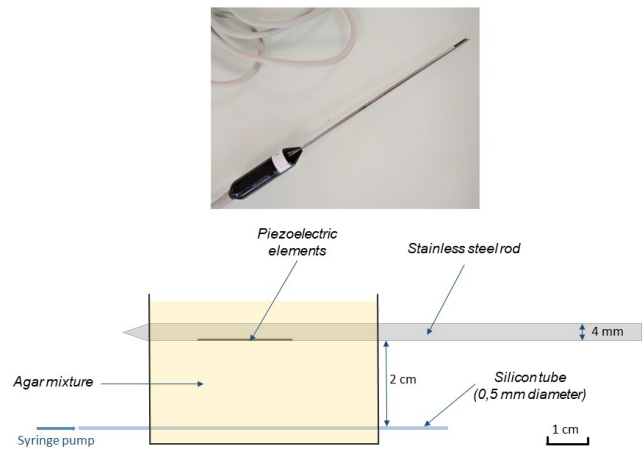

Fig. 1. Top: Picture of the probe. Bottom: Scheme of the probe inserted in the agar phantom.

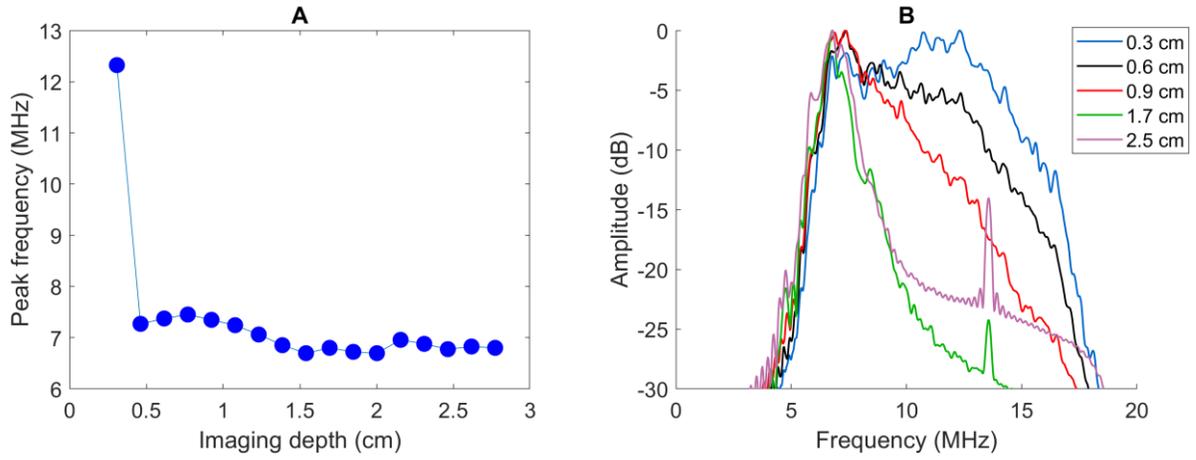

Fig. 2. (A) – Evolution of the peak frequency of the received spectrum as a function of imaging depth. (B) Received spectra as a function of imaging depth.

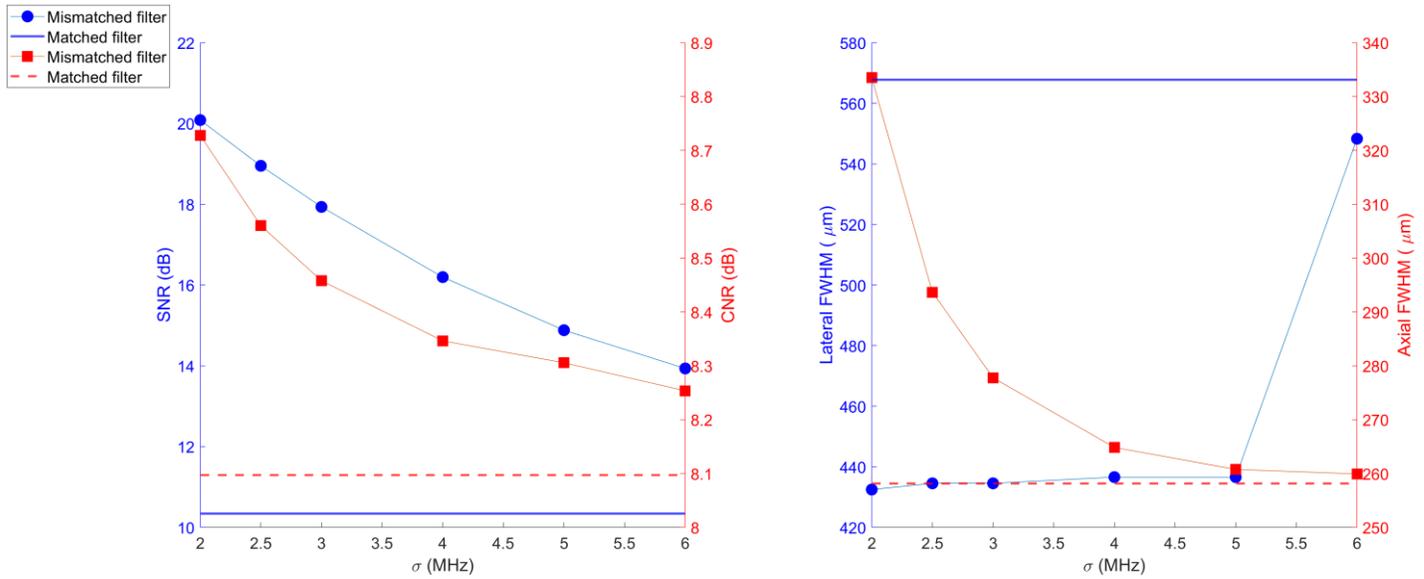

Fig. 3. (A) Evolution of the SNR and CNR measured in a chirp Power Doppler image as a function of the parameter $\sigma$ used to design the mismatched filter. (B) Lateral and axial FWHM for a chirp sequence as a function of the parameter $\sigma$ used to design the mismatched filter.

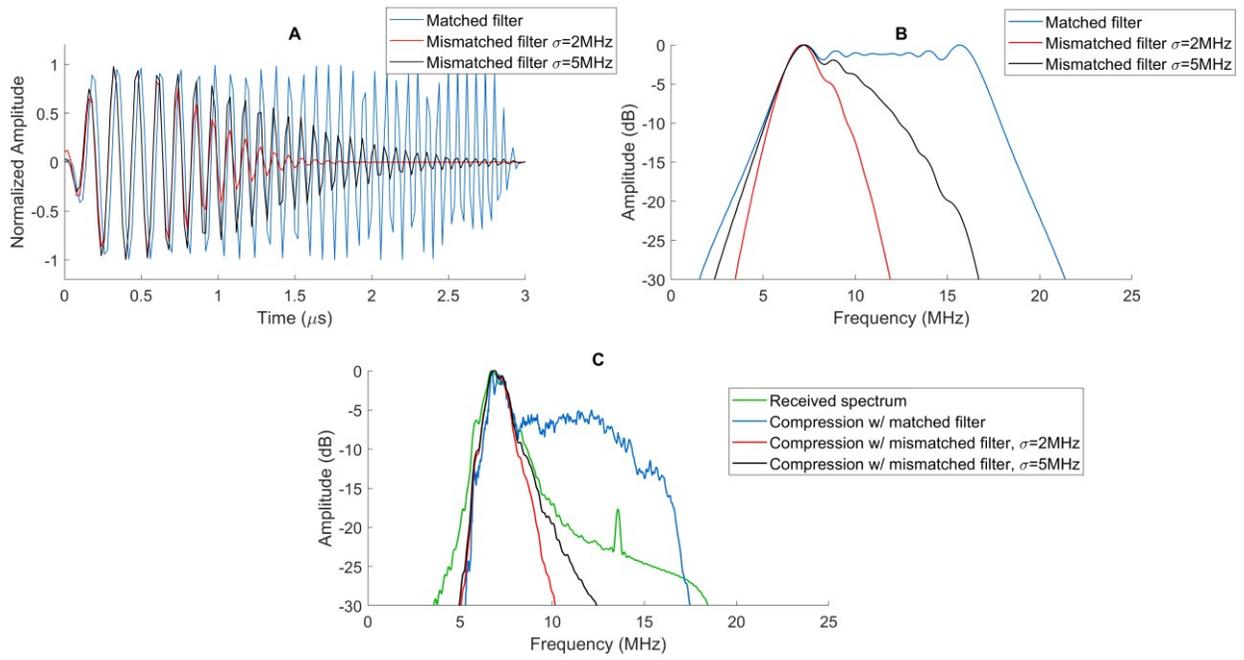

Fig. 4. (A) Temporal waveforms of the matched filter, the mismatched filter with $\sigma = 2$ MHz and the mismatched filter with $\sigma = 5$ MHz. (B) Spectra of the matched filter, the mismatched filter with $\sigma = 2$ MHz and the mismatched filter with $\sigma = 5$ MHz. (C) Averaged received spectrum for the chirp sequence before compression and after compression with a matched filter, a mismatched filter with $\sigma = 2$ MHz and a mismatched filter with $\sigma = 5$ MHz. The spectra have been acquired in a ROI corresponding to the measurement tube.

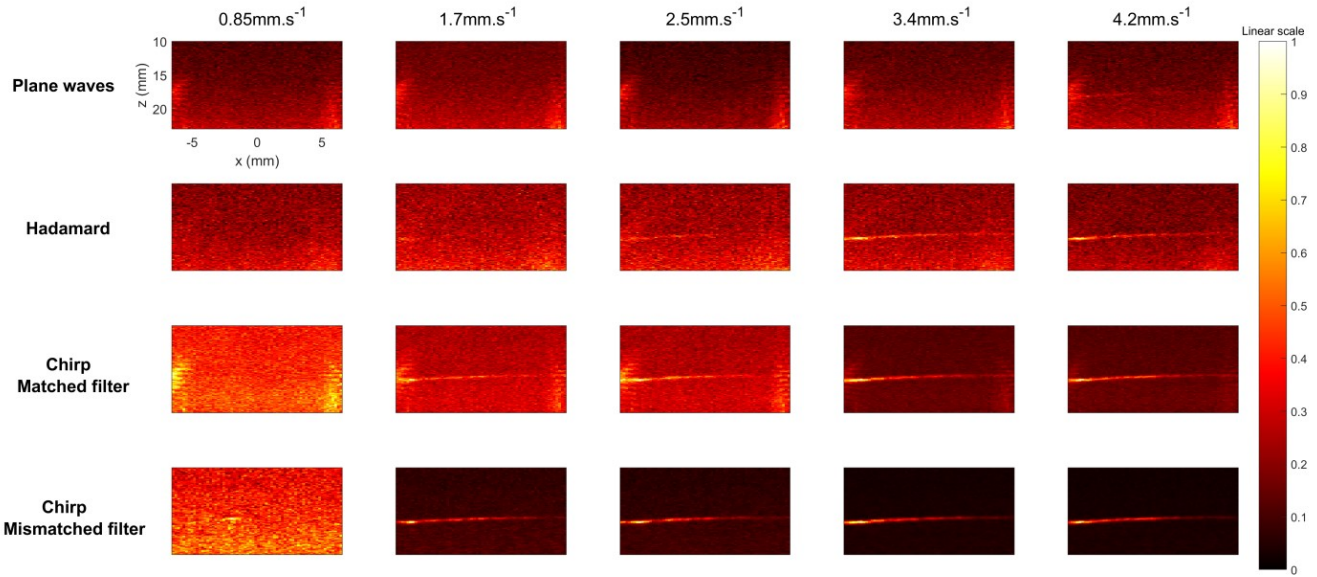

Fig. 5. (A) Evolution of power Doppler images as a function of the mean flow velocity for several emissions and compression methods: plane waves only, plane waves with Hadamard encoding for the emission, plane waves with chirp emission and a matched filter for the compression, plane waves with chirp emission and a mismatched filter with $\sigma = 2.5$ MHz for the compression.

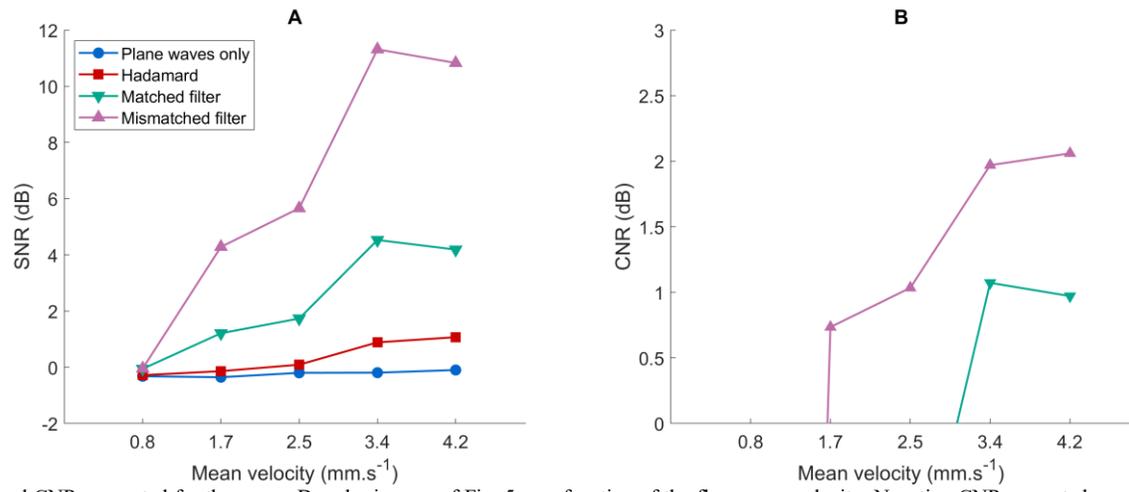

Fig. 6. SNR and CNR computed for the power Doppler images of Fig. 5 as a function of the flow mean velocity. Negative CNR are not shown for the sake of clarity.